\documentclass{ws-procs9x6}

\newcommand{\be}{\begin{equation}}
\newcommand{\ee}{\end{equation}}
\newcommand{\bea}{\begin{eqnarray}}
\newcommand{\eea}{\end{eqnarray}}
\def\nue{{\nu_e}}
\def\anue{{\bar\nu_e}}
\def\numu{{\nu_{\mu}}}
\def\anumu{{\bar\nu_{\mu}}}
\def\nutau{{\nu_{\tau}}}
\def\anutau{{\bar\nu_{\tau}}}
\newcommand{\dm}{\mbox{$\Delta{m}^{2}$~}}

\def\ltap{\ \raisebox{-.4ex}{\rlap{$\sim$}} \raisebox{.4ex}{$<$}\ }

\def\nue{{\nu_e}}
\def\anue{{\bar\nu_e}}
\def\numu{{\nu_{\mu}}}
\def\anumu{{\bar\nu_{\mu}}}
\def\nutau{{\nu_{\tau}}}
\def\anutau{{\bar\nu_{\tau}}}
\def\nul{{\nu_L}}
\def\anul{{\bar\nu_L}}

\begin{document}

\title{Electrophobic 
Lorentz invariance violation for neutrinos
and the see-saw mechanism}

\author{S.F.\ KING}

\address{Department of Physics and Astronomy, University of Southampton, \\
Highfield, Southampton S017 1BJ, UK\\
E-mail: sfk@hep.phys.soton.ac.uk}

\maketitle

\abstracts{
In this talk we show how Lorentz invariance
violation (LIV) can occur for Majorana neutrinos, without inducing
LIV in the charged leptons via radiative
corrections. Such ``electrophobic'' LIV is
due to the Majorana nature of the 
LIV operator together with electric charge conservation.
Being free from the strong constraints coming from the charged lepton sector,
electrophobic LIV
can in principle be as large as current neutrino experiments permit.
On the other hand electrophobic LIV could be naturally small 
if it originates from LIV in some 
singlet ``right-handed neutrino'' sector, and is felt in the physical
left-handed neutrinos via a see-saw mechanism. }

\section{Introduction}

In this talk we discuss a LIV scenario discussed in \cite{Choubey:2003ke}
with two desirable features:

{ (i) natural explanation of smallness of LIV

(ii) protection of LIV
in the neutrino sector from the 
bounds coming from the charged lepton sector}

We satisfy (i) by supposing that such effects originate
in the ``right-handed neutrino'' singlet sector,
and are only fed down to the left-handed neutrino sector via the
see-saw mechanism, thereby giving naturally small LIV in the 
left-handed neutrino sector. 

We satisfy (ii) by proposing a LIV operator which violates
lepton number by two units - forbidden by electric charge
conservation for charged fermions: ``electrophobic LIV''

The motivation for LIV in the right-handed neutrino sector is:

\begin{itemize}
\item 
Theoretically
attractive since  ``right-handed neutrinos'' 
could represent any singlet sector,
and need not be associated with ordinary quarks and leptons,
except via their Yukawa couplings to left-handed neutrinos. 

\item 
The
fact that LIV is associated only with such a singlet
sector could provide a natural explanation for why LIV appears to
be a good symmetry for charged fermions, while being potentially
badly broken in the neutrino sector. 

\end{itemize}

\section{CPTV in the right-handed neutrino sector}

Suppose that CPTV originates solely from the right-handed
sector due to the operator:
\be
\bar{N}_R^{\alpha}{B'}^\mu_{\alpha \beta}\gamma^\mu N_R^\beta
\ee

\begin{figure}[ht]
\centerline{\epsfxsize=4.1in\epsfbox{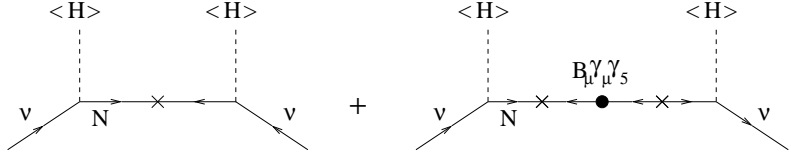}}   
\caption{See-saw mechanism with CPT violation in the right-handed
neutrino sector. \label{fig1}}
\end{figure}

The see-saw mechanism depicted in Fig.\ref{fig1}
leads to a naturally suppressed
CPT violating operator
in the left-handed neutrino sector: \cite{Mocioiu:2002pz}

\be
\bar{\nu}_L^{\alpha}b^\mu_{\alpha \beta}\gamma^\mu
\nu_L^\beta; \ \ \ \ b^\mu=\frac{m_{LR}^2{B'}^\mu}{({B'}^2+M_{RR}^2)}
\ee

\begin{figure}[ht]
\centerline{\epsfxsize=4.1in\epsfbox{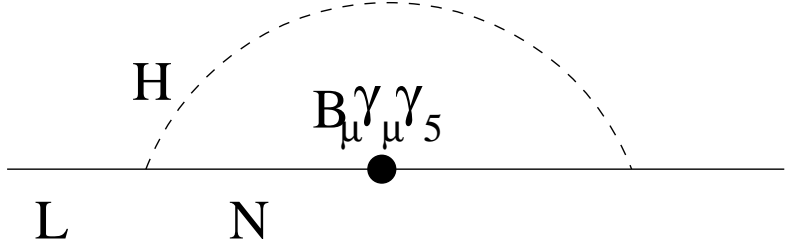}}   
\caption{One-loop contribution of CPT violation in the
right-handed neutrino sector to CPT violation in the charged lepton
sector. \label{fig2}}
\end{figure}

Mocioiu and Pospelov \cite{Mocioiu:2002pz} noted the following
problem, namely that
CPT violation is generated in the {\em charged} lepton sector via
one-loop radiative corrections as shown in Fig.\ref{fig2}.

The operator which is generated from Fig.\ref{fig2} is given by:
\be
\bar{L}_L^{\alpha}{b_{loop}}^\mu_{\alpha \beta}\gamma^\mu
L_L^\beta; \ \ L_L=(\nu_L ~e_L)^T
\label{L}
\ee
The CPT violating coefficient from Eq.\ref{L} is given by:
\be
b_{electron}\sim b_{loop}^\mu\sim
10^{-2}b^\mu
\ee
The electron CPTV limit in this coefficient is given by:
$b_{electron}<10^{-28}$ GeV which impllies that $b<10^{-26}$ GeV.

Is such a small amount of CPTV observable in the neutrino sector?
To answer this question, consider the 
constraints arising from the CPTV operator:
\be
 \anul^\alpha b_{\alpha \beta}^\mu
\gamma_\mu\nul^\beta  
\ee
It is conventional to consider the time component only
of this operator: 
\be
\anul^\alpha b_{\alpha \beta}^0
\gamma_0\nul^\beta 
\ee

The resulting two neutrino flavour equation of motion 
in the presence of CPTV is:
\be
i\frac{d}{dt}
\left(
\begin{array}{l}
\nue \\
\numu
\end{array}
\right)
= 
\left[A\left(
\begin{array}{ll}
-\cos2\theta & \sin2\theta \cr 
\sin2\theta & \cos2\theta
\end{array}\right)
+   B 
\left(\begin{array}{ll}
-\cos2\theta_b &
\sin2\theta_b \cr 
\sin2\theta_b & \cos2\theta_b
\end{array}\right)
\right ]
\left(
\begin{array}{l}
\nue \\
\numu
\end{array}
\right)
\ee
where
\be
A=\frac{\Delta m^2}{4E}, \ \ B=\frac{b^0_2-b^0_1}{2}
\ee
This results in the oscillation probability that an electron 
neutrino remains an electron neutrino given by:
\be
P_{ee}=1-\frac{D^2}{C^2+D^2}\sin^2\left(\sqrt{C^2+D^2}\ L\right)
\ee
where
\be
C=A\cos2\theta + B\cos2\theta_b; \ \ 
D=A\sin2\theta + B\sin2\theta_b
\ee
Neutrino oscillations are sensitive to $b\sim 10^{-20}$~GeV.
We therefore conclude that the electron CPTV limit $b_{electron}<10^{-28}$ GeV 
above renders any CPT violation in the
neutrino sector unobservable.

\section{Electrophobic LIV in the Right-Handed Neutrino Sector}
In order to overcome this problem we suggested the following 
LIV operator in the right-handed neutrino sector:\cite{Choubey:2003ke}
\be
{H'}_{\alpha\beta}^{\mu\nu}(\overline{N_R^C})_\alpha \sigma_{\mu\nu}
(N_R)_\beta; \ \ \Delta L=2
\label{M1}
\ee
\begin{figure}[ht]
\centerline{\epsfxsize=4.1in\epsfbox{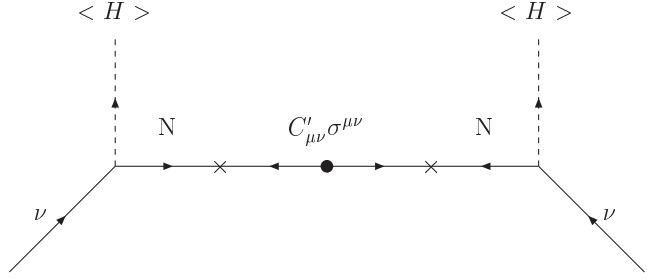}}   
\caption{See-saw contribution of LIV operator in the 
right-handed neutrino sector. \label{fig3}}
\end{figure}
The see-saw mechanism depicted in Fig.\ref{fig3} then
leads to naturally suppressed
LIV in the left-handed neutrino sector:
\footnote{This operator is reminiscent of the magnetic moment operator
$\mu_{\alpha\beta}(\overline{\nu_L^C})_\alpha \sigma_{\mu\nu}
(\nu_L)_\beta F^{\mu\nu}$. The main physical difference is that
our operator is independent of any physical magnetic fields, and can in
principle be arbitarily large.}
\be
h_{\alpha\beta}^{\mu\nu}(\overline{\nu_L^C})_\alpha \sigma_{\mu\nu}
(\nu_L)_\beta; \ \ 
h^{\mu\nu}=\frac{m_{LR}^2{H'}^{\mu\nu}}{({H'}^2+M_{RR}^2)}
\label{M2}
\ee

Note that both operators in Eqs.\ref{M1},\ref{M2} are Majorana operators.
They can never 
lead to LIV in the charged lepton sector to all orders of perturbation
theory due to electric charge conservation!

Expanding the electrophobic LIV operator in Eq.\ref{M2}:
\bea
h_{\alpha\beta}^{\mu\nu}(\overline{\nu_L^C})_\alpha \sigma_{\mu\nu}
(\nu_L)_\beta & = & 
 (\nu^C_{\alpha R})^\dagger \nu_{\beta L} H_+
 -  (\nu_{\alpha L})^\dagger \nu_{\beta R}^C H_- 
\nonumber \\
& + &  (\nu_{\beta L})^\dagger \nu_{\alpha R}^C H_- 
- (\nu_{\beta R}^C)^\dagger \nu_{\alpha L}H_+
\label{M3}
\eea
where $H_\pm = (h_{23} + h_{01}) \pm i(h_{13}
+h_{02})$.
Eq.\ref{M3} shows that electrophobic
LIV allows $\nu_\alpha \rightarrow \bar{\nu}_\beta$,
whereas the CPT considered previously
forbids $\nu_\alpha \rightarrow \bar{\nu}_\alpha$.

We now consider constraints on 
the coefficient which controls electrophobic LIV:
\be
H_\pm \rightarrow  h_{23} + h_{01} \equiv H
\ee
The two neutrino equation of motion is:
\be
i\frac{d}{dt}
\left(
\begin{array}{l}
\nu_{\alpha L} \cr 
\bar\nu_{\alpha R} \cr \nu_{\beta L} \cr \bar\nu_{\beta R} \cr
\end{array}
\right)
= 
\left(
\begin{array}{cccc}
-A \cos2\theta & 0 & A \sin2\theta & B \cr
0 & -A \cos2\theta & -B & A \sin2\theta \cr
A  \sin2\theta & -B & A  \cos2\theta & 0 \cr
B & A  \sin2\theta & 0 & A  \cos2\theta \cr
\end{array}
\right)
\left(
\begin{array}{l}
\nu_{\alpha L} \cr 
\bar\nu_{\alpha R} \cr \nu_{\beta L} \cr \bar\nu_{\beta R} \cr
\end{array}
\right)
\ee
where
\be
A=\frac{\Delta m^2}{4E}, \ \ B=H_{\alpha \beta}
\ee
This leads to the two-flavour oscillation probabilities:
\be
P_{\alpha \beta}= \frac{A^2\sin^22\theta}{A^2+B^2}
\sin^2\left(\sqrt{A^2+B^2}\ L   \right)
\ee
\be
P_{\alpha \bar \beta}
=
\frac{B^2}{A^2+B^2}
\sin^2\left(\sqrt{A^2+B^2}\ L   \right)
\ee
\be
P_{\alpha\alpha}=P_{\bar \alpha \bar \alpha}=
1-P_{\alpha \beta}-P_{\alpha \bar \beta}
\ee
\be
P_{\alpha \bar \alpha} =0
\ee

We now summarise the experimental constraints on electrophobic LIV
from different experiments.

{\bf Constraints from CHOOZ/Palo Verde:}

{CHOOZ and Palo Verde short baseline reactor experiments are 
consistent with no observed oscillation of $\bar{\nu_e}$ at 
baseline $L\sim 1$ km . This non-observation of any 
oscillations can be used to constrain 
{ $H_{e \bar\beta} \ltap 10^{-19}$ GeV}

[$H_{e \bar\beta}~(=H_{\bar e \beta}$ due to CPT invariance) 
is the LIV coeffecient responsible for 
$\anue(\nue) \rightarrow \nu_\beta(\nu_{\bar\beta})$ transition.}]

{\bf Constraints from the KamLAND experiment:}

{ KamLAND observes the electron antineutrinos produced in nuclear 
reactors from all over Japan and Korea. 
KamLAND results 
show a deficit of the antineutrino flux and are consistent with 
oscillations with \dm and mixing given by LMA solar solution.

KamLAND being a disappearance experiment is 
insensitive to whether the $\anue$ oscillate into $\numu$ 
due to mass and mixing or $\anumu$ due to LIV. 
%Even though the current KamLAND data, 
%has a strong evidence for suppression 
%of the incident antineutrino flux, the evidence for energy distortion 
%of the resultant spectrum is not very strong -- no distortion of the 
%spectrum is allowed at the 53\% C.L.. 
However LIV 
driven oscillations are inconsistent with the KamLAND 
energy distortion data leading to  
{ $H_{e \bar\beta} < 7.2 \times 10^{-22}$ GeV. }}
%Though this LIV solution is not as good as 
%oscillations with parameters in the 
%LMA region, it is still allowed 
%by the first results from the KamLAND experiment.
%It could be ruled out 
%and constraints tighter than those obtained from the 
%atmospheric neutrino data could be put on $b$, 
%if the future KamLAND data is consistent 
%$with spectral distortion.

{\bf Constraints from the atmospheric neutrino data:}

{ The atmospheric neutrino experiments observe a deficit of the 
$\numu$ and $\anumu$ type neutrinos, while the observed $\nue$ and $\anue$ 
are almost consistent with the atmospheric flux predictions.

The LIV term 
would convert $\numu$($\anumu$) into $\anutau$($\nutau$), 
while flavor oscillations 
convert $\numu$($\anumu$) to $\nutau$($\anutau$).

Since the experiments are insensitive to either $\nutau$ or $\anutau$, 
they will be unable to distinguish between the two cases.

LIV case 
is independent of the neutrino energy (same predicted 
suppression for the sub-GeV, multi-GeV, and the upward 
muon data). 
Therefore pure LIV term fails to explain the data but can 
exist as subdominant effect along with mass driven flavor 
oscillations, leading to limit: 
{ $H_{\mu \bar\tau} \ltap 10^{-20}$ GeV.}}

{\bf Constraints from the future long baseline experiments:}
{ Better constraints on LIV coefficient requires 
experiments with longer baselines.
MINOS and  
CERN to Gran Sasso (CNGS) experiments, ICARUS and OPERA,
have a baseline of about 732 km, though the  
energy of the $\numu$ beam in  
MINOS will be different from the energy of the CERN $\numu$
beam. However, since the LIV driven probability is independent 
of the neutrino energy, all these experiment would be expected 
to constrain
{ $H_{\mu \bar\beta} \ltap 10^{-22}$ GeV.} 

JPARC has 
shorter baseline of about 300 km only, while the NuMI off-axis 
experiment is expected to have a baseline not very different 
from that in MINOS and CNGS experiments.
The best 
constraints in terrestrial experiments 
would come from the proposed neutrino factory experiments, using 
very high intensity neutrino beams propagating over very large 
distances. 
Severe constraints, up to $H_{\mu \bar\beta} \ltap 10^{-23}$ 
GeV could be imposed for baselines of $\sim 10,000$ km.}

{\bf  Constraints from solar neutrinos:}

{ Neutrinos coming from the sun,
travel over very long baselines $\sim 1.5\times 10^8$ km.
So one could put stringent constraints on 
$H_{e \bar\beta}$ from the solar neutrino data.
However the situation for solar neutrinos is complicated 
due to the presence of large matter effects in the sun.}

{\bf Constraints from supernova neutrinos:}

{ Supernova are one of 
the largest source of astrophysical neutrinos, releasing 
about $3\times 10^{53}$ ergs of energy in neutrinos.
The neutrinos 
observed from SN1987A,
in the Large Magellanic Cloud, had traveled $\sim 50$ kpc to 
reach the earth.
Neutrinos from a supernova in our own galactic center 
would travel distances $\sim 10$ kpc. These would produce 
large number of events in the terrestrial detectors like 
the Super-Kamiokande. The observed flux and the energy 
distribution of the signal can then be used to 
constrain the LIV coefficient.}

{\bf  Constraints using the time of flight delay technique:}

{The violation of Lorentz invariance could also change the 
%dispersion relation and therefore the 
speed of the neutrinos and hence cause delay in their time of flight.
%Thus the neutrinos would come delayed with respect to 
%standard model particles with the same mass. 
The idea is to 
find the dispersion relation for the neutrinos in the presence 
of LIV and extract their velocity $v=\partial E/\partial p$, where 
$E$ is the energy and $p$ the momentum of the neutrino beam.
Then by comparing the time of flight of the LIV neutrinos, 
with particles conserving Lorentz invariance, one could 
{\it in principle} constrain the LIV coefficient.
The presence of the LIV term in the Lagrangian gives a 
see-saw suppressed correction to the mass term. Therefore
\be
v \approx 1 - \frac{m^2 + m^2_{LIV}}{E^2}
\nonumber
\ee
where $m$ is the usual mass of the neutrino concerned and 
$ m^2_{LIV}$ is the LIV correction. }

\section{Conclusion}

\begin{itemize}

\item
{ LIV may be introduced into a ``right-handed neutrino''
sector at some high scale, resulting in suppressed
LIV in the left-handed neutrino 
sector via the see-saw mechanism.}

\item
{ The $\Delta L=2$ lepton number violating 
operators induce LIV into the left-handed Majorana neutrino sector, 
while protecting LIV in the charged lepton sector to all orders
of perturbation theory due to electric charge 
conservation}

\end{itemize}

\end{document}